\documentclass[aps,prl,groupedaddress,twocolumn,showpacs]{revtex4-1}
\usepackage{graphicx,psfrag,amsmath,calc,amssymb, comment, color}

\begin{document}

\title{Tighter upper bounds on the critical temperature of two-dimensional \\
superconductors and superfluids: Approaching the supremum}

\author{Tingting Shi$^{1,2}$}
\author{Wei Zhang$^{1}$}
\author{C. A. R. S{\' a} de Melo$^{2}$}
\affiliation{$^{1}$ Department of Physics, Renmin University of China, Beijing 100872, China}
\affiliation{$^{2}$ School of Physics, Georgia Institute of Technology,
Atlanta, Georgia 30332, USA}

\date{\today}

\begin{abstract}
  We discuss standard and tighter upper bounds on
  the critical temperature $T_c$ of two-dimensional superconductors
  and superfluids versus particle density $n$ or filling factor $\nu$,
  under the assumption that the transition from the normal to the superconducting
  (superfluid) phase is governed by the
  Berezinskii-Kosterlitz-Thouless (BKT) mechanism of vortex-antivortex binding and 
  a direct relation between the superfluid density tensor and $T_c$ exists.
  The standard critical temperature upper bound
 $T_c^{up1}$ is obtained from the Glover-Ferrell-Tinkham sum rule for
  the optical conductivity, which constrains the superfluid density tensor components.
  However, we show that $T_c^{up1}$ is only useful in the limit of low particle/carrier density,
  where it may be close to the critical temperature supremum $T_c^{sup}$.
  For intermediate and high particle/carrier densities, 
  $T_c^{up1}$ is far beyond $T_c^{sup}$ for any given interaction strength.
  We demonstrate that it is imperative to consider at least the full effect of
  phase fluctuations of the order parameter for superconductivity (superfluidity)
  to establish tighter bounds over a wide range of densities.
  Using the renormalization group, we obtain the critical temperature supremum
  for phase fluctuations $T_c^{\theta}$ and show that it is a much tighter upper bound
  to $T_c^{sup}$ than $T_c^{up1}$ for all
  particle/carrier densities. We conclude by indicating that if the $T_c^{\theta}$ is exceeded
  in experiments involving single band systems, then a non-BKT mechanism must be invoked.
\end{abstract}
\maketitle

{\it Introduction:}
Several recent experiments have studied the critical temperature $T_c$ of 
two-dimensional (2D) superconductors as a function of carrier density $n$ or
filling factor $\nu$ for various materials, including double- and
triple-layered twisted graphene~\cite{herrero-2018, herrero-2021}, lithium-intercalated
nitrides~\cite{iwasa-2018, iwasa-2021} and sulphur-doped iron
selenide~\cite{shibauchi-2021}. In all these 2D systems, the
authors~\cite{herrero-2018, herrero-2021, iwasa-2018, iwasa-2021, shibauchi-2021}
describe their results as evolving from the Bardeen-Cooper-Schrieffer to the Bose regime
as $n$ or $\nu$ are changed from high to low, and have raised the issue
of the existence of an upper bound on $T_c$. For one-band 2D systems with
parabolic dispersion, the standard upper bound is known
to be $T_c^{up1} = \varepsilon_F/8$~\cite{botelho-2006, sharapov-1999}, in units where $k_B = 1$,
with $\varepsilon_F$ being the Fermi energy.
Extensions of this result have been proposed to flat band and
multiband systems~\cite{randeria-2019}. However, the validity of such extensions 
has been questioned in recent work showing several counter examples 
where upper bounds are arbitrarily exceeded~\cite{kivelson-2021}.

The question of the existence of an upper bound for $T_c$ in superconductors and
superfluids is of fundamental importance~\cite{kivelson-2018, kivelson-1995}.
Understanding the conditions under which such bounds exist for various systems
is key to paving the way to designing materials where room temperature
superconductivity can be achieved at ambient pressure, as suggested
by measurements of the penetration
depth of a variety of materials~\cite{uemura-1989, uemura-1991}.
However, upper bounds are practically useless if they are too far above the
supremum (least upper bound)~\cite{footnote-0},
thus when these upper bounds exist, it is essential to establish if they are tight,
that is, if they are close to the supremum $T_c^{sup}$.
Identifying tight upper bounds to $T_c^{sup}$ is a much more
difficult than merely determining a standard upper bound based on the
kinetic energy~\cite{randeria-2019}, nevertheless 
this is precisely what we propose to describe next.

Here, we study two examples where tighter upper bounds on $T_c$ 
can be established for one-band 2D systems: the continuum limit with parabolic
dispersion, and the square lattice case with cosinusoidal dispersion, both with
spatially dependent but non-retarded interactions.
We show that the standard upper bound obtained via the bare superfluid density
$\rho_s$~\cite{botelho-2006, randeria-2019}, which is independent of interactions
or symmetry of the order parameter, are practically useless
away from the regime of ultralow carrier density $n$ in the continuum or away from the
limits of $\nu \to 0$ or $\nu \to 2$ in the square lattice, because they severely overestimate
the supremum $T_c^{sup}$.
To remedy this issue, we demonstrate 
that much tighter bounds can be obtained by investigating the renormalized
superfluid density $\rho_s^R$ rather than the bare superfluid density $\rho_s$,
since $\rho_s^R \le \rho_s$ strictly holds. This relation arises physically because $\rho_s$ is
calculated in linear response theory, which does not include
the existence of vortices and antivortices in the superconductor or superfluid.
Large transverse current fluctuations,  due to vortices
and antivortices with quantized circulations,
screen the bare $\rho_s$ and renormalize it to $\rho_s^R$.
Furthermore, we show that the phase fluctuation supremum
$T_c^\theta = \pi \rho_s^R/2$ as a function of $n$ or $\nu$ and establish
that the supremum $T_c^{sup}$ must be always lower or
equal to $T_c^\theta$, that is $T_c^{sup} \le T_c^\theta$.
We also emphasize that tighter upper bounds for $T_c$, based on $T_c^\theta$,
rely on the idea that the transition
from the superconductor or superfluid to the normal state is driven by the
Berezinskii-Kosterlitz-Thouless (BKT)~\cite{berezinskii-1970, kosterlitz-thouless-1972}
vortex-antivortex unbinding mechanism. Finally, we conclude that if an experimental $T_c$
exceeds the phase fluctuation supremum $T_c^\theta$ then either the chosen model does
not apply or a non-BKT mechanism for superconductivity and superfluidity must be
invoked when the model applies.

{\it Continuum and Lattice Hamiltonians:} To construct tighter upper bounds on $T_c$,
we discuss 2D continuum and lattice Hamiltonians. In the continuum,
we start from the Hamiltonian density 
$
{\cal H} ({\bf r})
=
{\cal H}_{\rm K} ({\bf r})
+
{\cal H}_{\rm I} ({\bf r}), 
$
for a single band system in units where $\hbar=k_B=1$. The
kinetic energy density is
${\cal H}_{\rm K} ({\bf r})
=
\sum_s
\psi^\dagger_{s} ({\bf r}) \left[  -\frac{\nabla^2}{2m} \right] \psi_s ({\bf r})$, 
and the interaction energy density is 
${\cal H}_{\rm I} ({\bf r})
=
\int d^2 {\bf r}^\prime
V ({\bf r}, {\bf r}^\prime)
\psi_{\uparrow}^\dagger ({\bf r}) \psi_{\downarrow}^\dagger ({\bf r}^\prime)
\psi_{\downarrow} ({\bf r}^\prime)  \psi_{\uparrow} ({\bf r})$, 
with
$V ({\bf r}, {\bf r}^\prime) = - V_s g (\vert {\bf r} - {\bf r}^\prime\vert/ R)$.
The  magnitude of the $s$-wave attractive interaction $V_s$ has units of energy,
and the dimensionless function $g (\vert {\bf r} - {\bf r}^\prime\vert/ R)$ has  spatial range $R$.
Fermions with spin projection $s$ at position ${\bf r}$ are represented by the 
the field operator $\psi_s^{\dagger} ({\bf r})$. 
In the square lattice, we start from an extended Fermi-Hubbard Hamiltonian 
$H =
-t \sum_{\langle ij \rangle, s}  \psi_{i s}^\dagger  \psi_{j s}
+
U \sum_{i s} {\hat n}_{i \uparrow} {\hat n}_{i \downarrow}
+
\sum_{i < j s s^\prime} V_{ij} {\hat n}_{i s} {\hat n}_{j s^\prime}$, 
where ${\hat n}_{i s} = \psi_{i s}^\dagger \psi_{j s}$ is the fermion number
operator at site $i$ with spin $s$. The nearest neighbor hopping is $t$,
the local (on-site) interaction is $U$,
and the interaction between fermions in sites $i$ and $j$ is $V_{ij}$.

{\it Hamiltonians in Momentum Space:} The Hamiltonians in momentum space, for both
continuum and lattice, are
\begin{equation}
\label{eqn:hamiltonian-momentum-space}
H =
 \sum_{{\bf k} s} 
\varepsilon_{\bf k} \psi_{{\bf k}, s}^\dagger \psi_{{\bf k}, s} + 
\sum_{{\bf k} {\bf k}^\prime {\bf q}} V_{{\bf k}{\bf k}^\prime}
b_{{\bf k}{\bf q}}^\dagger b_{{\bf k}^\prime {\bf q}} ,
\end{equation}
where 
$b_{{\bf k}{\bf q}} = \psi_{-{\bf k}+{\bf q}/2, \downarrow} \psi_{{\bf k}+{\bf q}/2, \uparrow}$
is the pairing operator, 
$\varepsilon_{\bf k} = {\bf k}^2/2m$ for the continuum, 
and $ \varepsilon_{\bf k} = - 2t \left[ \cos (k_x a) + \cos (k_y a) \right] $ for the square lattice.
The summation over ${\bf k}$ and ${\bf k}^{\prime}$ represent integrals
in the continuum and discrete sums in the lattice. The momentum-space interaction
$V_{{\bf k} {\bf k}^\prime}$ is the double Fourier transform of the real space interactions.

In the examples discussed below, we focus on $s$-wave pairing,
and thus, we will use an expansion of 
the momentum space interaction $V_{{\bf k} {\bf k}^\prime}$ in terms of its continuum
or lattice angular momentum~\cite{duncan-2000, annett-1990, iskin-2005}
and keep only the $s$-wave component. In this case,
the interaction potential can be approximated by the separable form
\begin{equation}
\label{potential}
V_{{\bf k}{\bf k}'} = -V_s 
\Gamma_s ({\bf k}) \Gamma_s({\bf k}^{\prime}) .
\end{equation}
In the continuum, for an attractive well with depth $V_s$ and radius $R$,
the symmetry factor can be approximated by 
$
\Gamma_s ({\bf k}) = 
\left( 1 + k/k_R \right)^{-1/2},
$
where $k_R \sim  R^{-1}$ plays the role of the interaction range in momentum
space~\cite{footnote-1}. In the square lattice, which has $C_4$ point group, the
symmetry factor for conventional $s$-wave pairing is $\Gamma_s ({\bf k}) = 1$,
when, for instance,  only local attractive interactions are considered,
while the symetry factor for extended $s$-wave pairing is
$\Gamma_s ({\bf k}) = \cos (k_x a) + \cos (k_y a)$, when, for example, nearest neighbor
attractive interactions are included.
For more general lattices, $\Gamma_s ({\bf k})$ are given by
the irreducible representations of the point group of the lattice~\cite{annett-1990}
compatible with $s$-wave symmetry. 

{\it Effective Action:} We introduce the chemical potential $\mu$ and  the order parameter
for superfluidity $\Delta$ in terms of its modulus $\vert \Delta \vert$ and phase $\theta$.
The effective action is
$
S_{\rm eff} =
S_{\rm sp} \left( \vert \Delta \vert \right)
+ 
S_{\rm ph} ( \vert \Delta \vert, \theta).
$
The first contribution to $S_{\rm eff}$ is the 
saddle-point action 
$
S_{\rm sp} 
=
\sum_{\bf k}
\left[
\frac{\left( \xi_{\bf k} - E_{\bf k} \right)}{T}
-
2 \ln \left(  1 + e^{-E_{\bf k}/T}\right)
\right] + \frac{\vert \Delta \vert^2}{T V_s},
$
where
$
E_{\bf k} = \sqrt{\xi_{\bf k}^2 + |\Delta_{\bf k}|^2}
$
is the energy of quasiparticles with $\xi_{\bf k} = \varepsilon_{\bf k} - \mu$, and 
$\Delta_{\bf k} = \Delta \Gamma_{s} ({\bf k})$ is the order parameter
function for $s$-wave pairing.
The second contribution
\begin{equation}
\label{eqn:phase-fluctuation-action}
S_{\rm ph} 
= \frac {1} {2} \int dr 
\left\{
\sum_{ij} \rho_{ij} \, \partial_i \theta(r) \partial_j \theta(r)
+
\kappa_s \left[ \partial_{\tau} \theta (r) \right]^2 
\right\},
\end{equation}
represents the phase-only fluctuation action 
in the low frequency and long wavelength limit.
Here, the integrals run over position and imaginary time $r = ({\bf r}, \tau)$, with
$\int dr \equiv \int_0^{1/T} d\tau \int d^2{\bf r}$.
In Eq.~(\ref{eqn:phase-fluctuation-action}), 
\begin{equation}
\label{eqn:superfluid-density}
\rho_{ij} 
=
\frac {1} {4 L^2} \sum_{\bf k}
\left[
2 n_{\rm sp} ({\bf k}) \partial_i \partial_j \xi_{\bf k} - 
Y_{\bf k} \partial_i \xi_{\bf k} \partial_j \xi_{\bf k}
\right] ,
\end{equation}
is the superfluid density tensor,  where $\partial_i$ is the partial 
derivative with respect to momentum $k_i$ with $i = \{ x, y \}$.
Here,
$n_{\rm sp} ({\bf k}) = \frac {1} {2} \left[
1 - (\xi_{\bf k}/ E_{\bf k}) 
\tanh \left(  E_{\bf k} / 2 T \right) \right]$ 
is the momentum distribution per spin state, and
$
Y_{\bf k} = (2T)^{-1} 
{\rm sech}^2 (E_{\bf k} / 2T)
$
is the Yoshida function.
The superfluid density tensor is diagonal
$\rho_{ij} = \rho_s \delta_{ij}$, where
$\rho_s = (1/4mL^2) \sum_{\bf k} \left[ 2 n_{sp} ({\bf k}) - (k_x^2/m) Y_{\bf k}\right]$
for the continuum and
$\rho_s =
(ta^2/L^2) \sum_{\bf k} \left[  \cos (k_x a) n_{sp}({\bf k}) - t \sin^2 (k_x a) Y_{\bf k} \right]$
for the square lattice.
The second term in Eq.~(\ref{eqn:phase-fluctuation-action}) is
\begin{equation}
\label{eqn:compressibiity}
\kappa_{s} 
=
\frac {1} {4 L^2} \sum_{\bf k} 
\left[
\frac {|\Delta_{\bf k}|^2} { E_{\bf k}^3 }
\tanh \left( \frac { E_{\bf k} }  { 2 T}  \right) + 
\frac { \xi_{\bf k}^2 } { E_{\bf k}^2 } Y_{\bf k}
\right] ,
\end{equation}
with $\kappa_s = \kappa/4$, where
$\kappa = \partial n/ \partial \mu \vert_{T,V}$ is related to the
thermodynamic compressibility ${\cal K} = \kappa/n^2$.

The phase of the order parameter can be separated as 
$\theta ({\bf r}, \tau) = \theta_c ({\bf r}, \tau) + \theta_v ({\bf r}) $,
where the $\tau$-dependent (quantum) term
$\theta_c ({\bf r}, \tau)$ is due to collective modes (longitudinal velocities),
and the $\tau$-independent (classical) term $\theta_v ({\bf r})$ is due to vortices 
(transverse velocities). This leads to the action $S_{\rm ph} = S_{c} + S_{v}$,
since the longitudinal and transverse velocities are orthogonal.
The collective mode action is 
$
S_{c} = \frac{1}{2} \int dr 
\left[
\rho_s \left[ \nabla \theta_{c} (r) \right]^2
+
\kappa_{s} \left[ \partial_\tau \theta_{c} (r) \right]^2
\right],
$
while the vortex action is  
$
S_{v}
=
\frac{1}{2T} \int d^2{\bf r} \,
\rho_s\left[ \nabla \theta_v ({\bf r}) \right]^2.
$
The vortex contribution arises from the transverse velocity
${\bf v}_t = \nabla \theta_v ({\bf r})$, where $\nabla \cdot {\bf v}_t ({\bf r}) = 0$,
by using the relation
$
\nabla \times {\bf v}_t ({\bf r})
=
2\pi {\hat {\bf z}} n_v ({\bf r})
$
where
$
n_v ({\bf r})
=
\sum_{i} n_i \delta ({\bf r} - {\bf r}_i ),
$
is the vortex density and $n_i = \pm 1$ is the vortex topological charge (vorticity)
at ${\bf r}_i$. Using these relations, we write
\begin{equation}
\label{eqn:vortex-action-core-energy}
S_{v}
=
2\pi \frac{\rho_s}{2T} \sum_{i \ne j} n_i n_j G ({\bf r}_i - {\bf r}_j)
+ \sum_i \frac{E_c}{T} n_i^2 ,
\end{equation}
where $E_c$ is the vortex core energy, and $G ({\bf r}_i - {\bf r}_j)$ is the
interaction potential between the topological charges $n_i$ and $n_j$
satisfiying Poisson's equation $\nabla^2_{\bf r} G ({\bf r} - {\bf r}^\prime) = 0.$ 

{\it Critical Temperature:} The self-consistency relations for $\vert \Delta \vert$
and $\mu$ for a given temperature $T$ are obtained from 
effective action $S_{\rm eff} = S_{\rm sp} + S_{\rm ph}$ as follows.
The order parameter equation is obtained through the stationarity condition 
$
\delta S_{\rm sp} / 
\delta \Delta^{*} = 0 ,
$
leading to 
\begin{equation}
\label{eqn:order-parameter}
\frac {1} {V_{s}}  = 
\sum_{\bf k} \frac {|\Gamma_{s} ({\bf k})|^2} {2 E_{\bf k}}
\tanh \left( \frac {E_{\bf k}} {2 T} \right) .
\end{equation}

To relate $\mu$ and $n = N/L^2$, where $N$ is the total number of particles per band
and $L^2$ is the area of the sample, we use the thermodynamic relation
$
n =
-\partial {\widetilde \Omega}/ \partial\mu \vert_{T, V},
$
with ${\widetilde  \Omega} = \Omega/L^2$, where
$\Omega$ is the thermodynamic potential~\cite{thermodynamic-potential}.
This leads to the number equation
$
n =  n_{\rm sp} + n_{\rm cm} + n_{\rm zp}.
$
Here, 
$
n_{j} = - \partial {\widetilde \Omega}_{j} / \partial \mu \vert_{T,V}$,
with $j = \{{\rm sp}, {\rm cm}, {\rm zp} \}$.
Here,  $n_{\rm sp}= 2 \sum_{\bf k} n_{\rm sp} ({\bf k})$, while
$n_{\rm cm}$ and $n_{\rm zp}$ are  obtained from their respective
${\widetilde\Omega}_j$~\cite{thermodynamic-potential}.

The critical temperature, within the BKT mechanism, is given by
the Nelson-Kosterlitz~\cite{nelson-kosterlitz-1977} relation 
\begin{equation}
\label{eqn:critical-temperature}
T_{c}^{\theta}
= \frac {\pi} {2} \rho_s^{R}  ( \mu, \vert \Delta \vert, T_c^{\theta} ),
\end{equation}
as $T \to T_c^{\theta}$ from below, 
where 
\begin{equation}
\label{eqn:renormalized-superfluid-density} 
\rho_s^{R}
=
\rho_s -
\frac{\rho_s^2}{2T}
\lim_{{\bf q} \to {\bf 0}}
\frac{\langle n_v ({\bf q}) n_v (-{\bf q}) \rangle}{\vert {\bf q} \vert^2}, 
\end{equation}
is the renormalized superfluid density.
Here, $n_v ({\bf q})$ is the Fourier transform of vortex density
$n_v ({\bf r})$.
From Eq.~(\ref{eqn:renormalized-superfluid-density}), it is clear
that $\rho_s^{R} \le \rho_s$ at any temperature $T$, because the correlation
function
$
{\cal F}
=
\lim_{{\bf q } \to {\bf 0}} \langle  n_v ({\bf q}) n_v (- {\bf q}) \rangle/\vert {\bf q} \vert^2
$
is strictly non-negative, that is, ${\cal F} \ge 0$.
This implies that the upper bound $T_c^{up1}$ based
on the bare superfluid density $\rho_s$~\cite{botelho-2006, randeria-2019} is not tight, and
therefore may severely overestimate the least upper bound, that is,
the supremum $T_c^{sup}$. The relation in
Eq.~(\ref{eqn:critical-temperature}) must be viewed as the phase fluctuation
supremum (least upper bound) $T_c^{\theta}$ and as tighter upper bound to
the supremum $T_c^{sup}$.

To establish $T_c^{\theta}$, we combine
Eqs.~(\ref{eqn:vortex-action-core-energy}) and~(\ref{eqn:renormalized-superfluid-density})
to obtain the solution of the renormalization group
flow equations~\cite{kosterlitz-1974, nelson-1977, chaikin-1995} leading to 
\begin{equation}
\label{eqn:RG-solution}
y^2(l) - \frac{1}{2\pi^3}
\left[
\frac{2}{K(l)} + \pi {\rm ln} K(l)
\right] =  A,
\end{equation}
for running variables $K(l)$ and $y(l)$.  The initial conditions are $K(0) = \rho_s/T$
and $y(0) = {\rm exp}(-E_c/T)$, satisfying  the relation
$y(0) = {\rm exp}{\left[ -E_c K (0)/\rho_s\right] }$. The flow of $\rho_s$
is described by $K (\ell) $, and
the flow of the vortex fugacity is represented by $y (\ell)$. 
\begin{figure}[t]
\begin{center}
\includegraphics[width=0.98\linewidth]{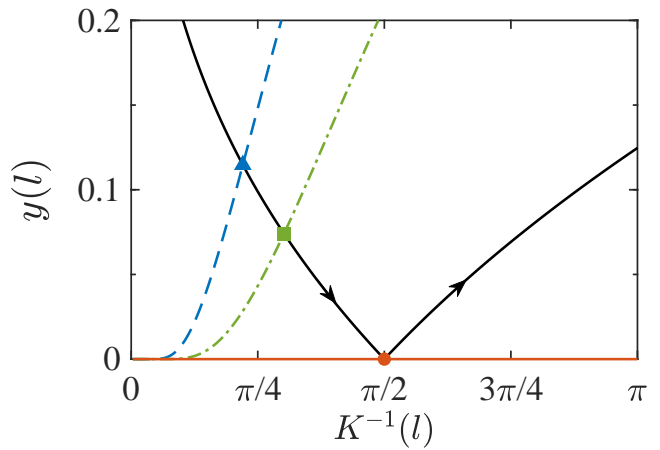}
\end{center}
\vspace{-6mm}
\caption{Critical flow line (solid black) and examples of initial conditions
$y(0) = e^{-E_c K(0)/\rho_s}$ lines for $E_c = 1.5 \rho_s$ (dashed blue),
for $E_c = \pi^2 \rho_s/4$ (dot-dashed green), and $E_c \to \infty$ (solid red).
The critical points are the solid blue triangle for $E_c = 1.5 \rho_s$,
the solid green square for $E_c = \pi^2 \rho_s/4$ (XY model), and
the solid red  circle for $E_c \to \infty$, which is also the fixed point of critical flow line.
When $E_c \to \infty$, $T_c^{\theta} \to \pi \rho_s/2$, thus for any  $E_c \le  \infty$, then
$T_c^{\theta} = \rho_s/K_c \le  \pi \rho_s/2$.
}
\label{fig:one}
\end{figure}
The constant $A$ determines the family of flow curves in the $K^{-1}$-$y$ plane.
In Fig.~\ref{fig:one}, the critical flow curve,  for which
$A = \left[ \ln \pi/2 - 1 \right]/\pi^2 = -0.0278$, is shown.
The fixed point at $(K_*^{-1}, y_*) = (\pi/2, 0)$ leads to 
the relation $T_c^\theta/\rho_s^R = \pi/2$ in Eq.~(\ref{eqn:critical-temperature}).
From the intersection between the curve
$y(0) = {\rm exp}{\left[ -E_c K (0)/\rho_s\right] }$ and the critical flow line, we obtain the
relation $T_c^\theta/\rho_s = K_c^{-1}$, which allows to relate $\rho_s^R$ and $\rho_s$ 
via $\rho_s^R = \rho_s (2/\pi K_c)$. As seen in Fig.~\ref{fig:one},  $K_c^{-1} \le \pi/2$,
that is, $( 2/\pi K_{c}) \le 1$ for any value
of the vortex core energy $E_c$. Therefore, $\rho_s^{R}$ is
always less or equal to $\rho_s$, that is, $\rho_s^{R} \le \rho_s$. The equality between
$\rho_s^R$ and $\rho_s$ occurs only when $E_c \to \infty$.

An important consequence of $\rho_s^R \le \rho_s$ is that
$T_c^\theta = \pi \rho_s^R/2 \le \pi \rho_s/2$. Since the supremum $T_c^{sup}$ is always
upper-bounded by the phase fluctuation supremum $T_c^\theta$, that is,
$T_c^{sup} \le T_c^{\theta}$,
it is clear that a lower upper bound is reached using $\rho_s^R$ rather than $\rho_s$.
Physically, $T_c^{\theta}$ is a better upper bound to $T_c^{sup}$ than
$T_c^{ub0} = \pi \rho_s (T = 0)/2$ as
suggested for superconductors with small superfluid density~\cite{kivelson-1995},
where phase fluctuations are important. Furthermore,
$T_c^{\theta}$ is also a better upper bound of $T_c^{sup}$
than $T_c^{ub1}= \pi \rho_{s_1}/2$, where
$\rho_{s_1} = (1/4L^2) \sum_{\bf k} 2 n_{sp} ({\bf k}) \partial_{x}^{2} \xi_{\bf k}$
is the first term of Eq.~(\ref{eqn:superfluid-density})  when $\rho_{ij} = \rho_s \delta_{ij}$
and $\xi_{\bf k}$ is isotropic in the continuum (${\rm C}_\infty$ or ${\rm SO}(2)$ symmetric)
or in the square lattice (${\rm C}_4$ symmetric).
In the continuum, $\rho_{s_1} \le  n/4m$~\cite{botelho-2006}
is upper bounded, at any temperature $T$, by the ratio
between the maximum pair density $n/2$ and the fermion pair mass $2m$ and
reflects the Ferrell-Glover-Tinkham
(FGT)~\cite{ferrell-1958, tinkham-1959, tinkham-1975} sum rule
for the optical conductivity, first derived by Kubo~\cite{kubo-1957}.
This leads to the standard upper bound
$T_c^{ub1} = \varepsilon_F/8$~\cite{botelho-2006, randeria-2019}, which is also
known to apply
to anisotropic superfluids~\cite{devreese-2014, devreese-2015, randeria-2019}, like
those with spin-orbit coupling~\cite{devreese-2014, devreese-2015}.
For the square lattice, $\rho_{s_1} = (t a^2/L^2) \sum_{\bf k} \cos (k_x a) n_{sp} ({\bf k})
\le t \nu/2$~\cite{footnote-lattice}, where $\nu$ is the filling factor of the band,
is also consistent with the FGT optical sum rule~\cite{hirsch-2000, chubukov-2011}.
Using particle-hole symmetry, a similar bound
$\rho_{s_1} \le t (2 - \nu)/2$ applies, leading to the standard upper bound
$T_c^{up1} = t (\pi/4) {\rm min} \{ \nu, (2 - \nu) \}$.
The sequence of temperatures is $T_c^{sup} \le T_c^{\theta} \le T_c^{up0} \le T_c^{up1}$,
with $T_c^{\theta}$ being the best upper bound to $T_c^{sup}$ and $T_c^{up1}$ being the worst.

\begin{figure}[t]
\begin{center}
\includegraphics[width=0.98\linewidth]{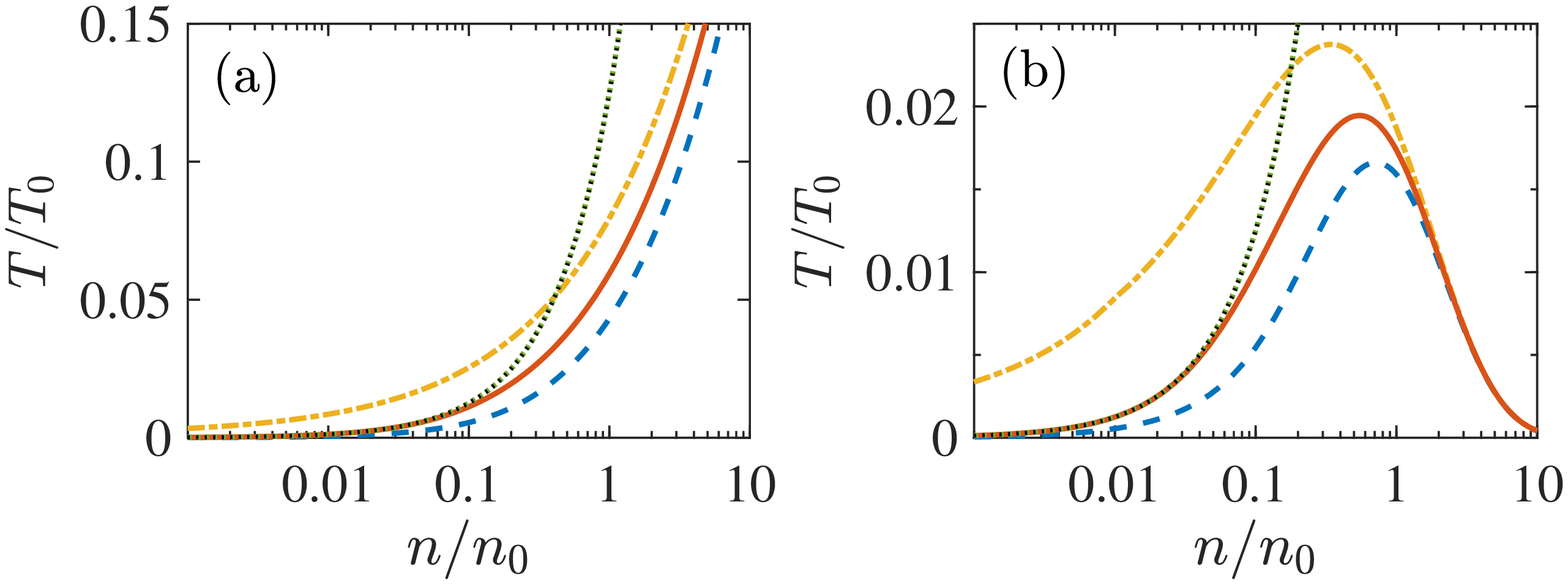}   
\end{center}
\vspace{-6mm}
\caption{Plots of $T_c$ versus $n$, in units of
$\varepsilon_0 = k_0^2/2m = T_0$ and $n_0 = k_0^2/2\pi$. 
Results for zero-ranged interactions $k_R \to \infty$ are shown in (a)
and for finite-ranged interactions with $k_R = k_0$ are shown in (b). 
In both panels the two-body binding energy
is $E_B = 0.01 \varepsilon_0$.
The dotted black lines are $T_c^{up1} = \varepsilon_F/8$,
the dotted green lines are  $T_c^{up0}= \pi \rho_s (T = 0)/2$, the
dot-dashed yellow lines are $T_{c}^{\rm mf}$, the solid red lines
are $T_c^{\theta}$ for $E_c \to \infty$, and the dashed blue lines are
$T_c^{\theta}$ for $E_c = 1.5 \rho_s$. 
}
\label{fig:two}
\end{figure}
{\it Results:} The phase fluctuation supremum $T_c^\theta$ is a much tighter
upper bound to the supremum $T_c^{sup}$ in comparison to both 
$T_c^{up1}$ based on $\rho_{s_1}$, and $T_c^{up0}$ based on $\rho_s (T = 0)$.
This also applies to the saddle point (mean field) critical temperature $T_c^{\rm mf}$
obtained by neglecting phase fluctuations.
The self-consistency relations determine $\mu$, $\vert \Delta \vert$ and $T_c^{\theta}$ as
functions of density $n$ in the continuum or filling factor $\nu$ in the square lattice for
given interaction parameters.

In Fig.~\ref{fig:two}, we show $T_c^\theta$, $T_c^{up0}$, $T_c^{up1}$, and $T_c^{\rm mf}$
versus $n$ for the cases of zero-ranged and finite ranged potentials.
We use temperature/energy 
$T_0 = \epsilon_0 = k_0^2/2m$ and density $n_0 = k_0^2/2\pi$ units, where $k_0$ is
a reference momentum related to the unit cell length $a$ of a crystal ($k_0 = 2\pi/a$) or 
to the laser wavelength $\lambda$ in a cold atom
system ($k_0 = 2\pi/\lambda$), in which case, $k_0$ $(\varepsilon_0)$ represents
the recoil momentum (energy). We convert the interaction $V_s$ with range $k_R$
into the two-body binding energy $E_B$~\cite{footonote-binding-energy}
to compare more easily the cases of zero and finite ranges.
In comparison to the phase fluctuation supremum $T_c^{\theta}$ (solid red line)
for $E_c \to \infty$,
the standard upper bound $T_c^{up1}$ (dotted black line)
fails miserably at intermediate and high densities $n$ 
being practically useless in that regime.
For parabolic bands,  $T_c^{up0}$ (dotted green line)  is equal to $T_c^{up1}$ for all
$n$ due to Galilean invariance. Furthermore, 
$T_c^{\rm mf}$ is always larger than  $T_c^{\theta}$,
exceeds $T_c^{up1} = T_c^{up0}$ at lower densities, and is only reliable at larger densities.

\begin{figure}[t]
\begin{center}
\includegraphics[width=0.98\linewidth]{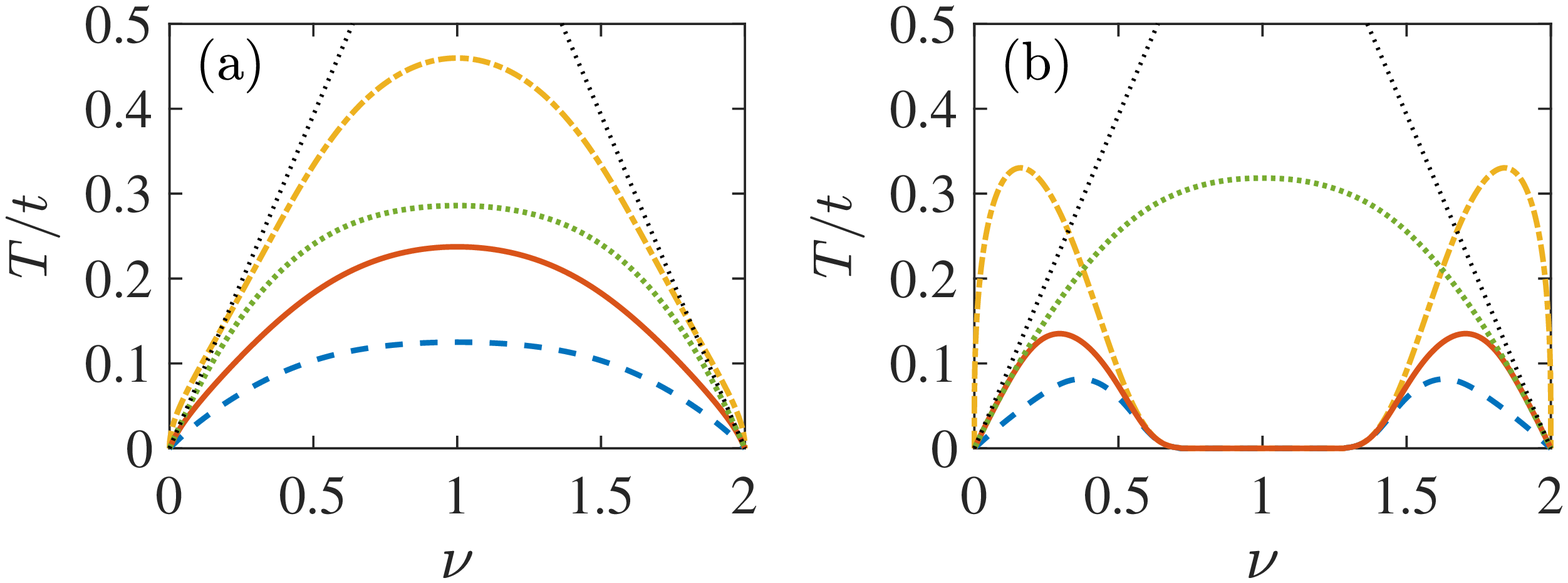}   
\end{center}
\vspace{-6mm}
\caption{
Plots of $T_c$ versus $\nu$. In both panels the interaction parameter is
$V_s/t  = 3$.  In (a) $\Gamma_s ({\bf k}) = 1$ (conventional s-wave), and
in (b) $\Gamma_s ({\bf k}) = \cos (k_x a) + \cos (k_y a)$ (extended s-wave).
The dotted black lines are the standard upper bound
$T_c^{up1} = t (\pi/4) {\rm min} \{ \nu, (2 - \nu) \}$,
the dotted green lines are $T_c^{up0} = \pi \rho_s (T =0)/2$, 
the dot-dashed yellow lines are $T_{c}^{\rm mf}$,
the solid red lines are $T_c^{\theta}$ for $E_c \to \infty$,
and the dashed blue lines are $T_c^{\theta}$ for $E_c = 1.5 \rho_s$. 
}
\label{fig:three}
\end{figure}

In Fig.~\ref{fig:three}, we show $T_c^\theta$, $T_c^{up0}$, $T_c^{up1}$, and $T_c^{\rm mf}$
versus $\nu$ for $V_s/t = 3$. In panel (a) $\Gamma_s ({\bf k}) = 1$ (conventional $s$-wave)
and in (b) $\Gamma_s ({\bf k}) = \cos (k_x a) + \cos (k_y a)$
(extended $s$-wave)~\cite{extended-hubbard-model}.
In comparison to the phase fluctuation supremum $T_c^{\theta}$ (solid red line)
for $E_c \to \infty$, the standard upper bound $T_c^{up1}$ (dotted black line)
fails miserably at intermediate fillings $\nu$, where even $T_c^{\rm mf}$ is a better
upper bound, however $T_c^{up1}$ is tighter for $\nu \sim 0$ or $\nu \sim 2$.
Notice that $T_c^{up0}$ is always a better upper bound than $T_c^{up1}$, and while in
(a) it is not too far above $T_c^{\theta}$, in (b) it overestimates substantially $T_c^{\theta}$
at intermediate $\nu$, where the order parameter modulus $\vert \Delta \vert$ vanishes.
In (a), $T_c^{\rm mf}$ is always above all upper bounds for $\nu \sim 0$ and
$\nu \sim 2$, but below $T_c^{up1}$ and above $T_c^{up0}$ and $T_c^{\theta}$ at
intermediate $\nu$. While in (b), $T_c^{\rm mf}$ is above all upper bounds for $\nu \sim 0$ or
$\nu \sim 2$, but at
intermediate values of $\nu$ it is below $T_c^{up1}$ and $T_c^{up0}$, but always
above $T_c^{\theta}$. Quantum Monte Carlo (QMC) data~\cite{scalettar-2004}
on $T_c$ versus $\nu$ for the attractive Hubbard model in (a) are bounded by $T_c^{\theta}$.

{\it Beyond Phase Fluctuations:} We remark that 
effects beyond phase fluctuations (longitudinal and tranverse), such as modulus
fluctuations of the order parameter, are not included in the phase fluctuation action
describing  the BKT mechanism for 2D superconductivity/superfluidity.
Modulus fluctuations may further renormalize the superfluid density $\rho_s$ and
the compressibility $\kappa_s$~\cite{footnote-fisher}, however these
effects can only reduce $T_c^{\theta}$. Thus, we can safely regard $T_c^{\theta}$
as the phase fluctuation supremum critical temperature for any given $E_c$.

{\it Conclusions:}
We investigated tighter upper bounds on the critical temperature of two-dimensional (2D)
superconductors and superfluids with a single parabolic (cosinusoidal) band in the continum
(square lattice). Using the renormalization group, we obtained the phase fluctuation
supremum critical temperature $T_c^{\theta}$  as the best upper bound for
the supremum $T_c^{sup}$ within the Berezinksii-Kosterlitz-Thouless (BKT)
vortex-antivortex binding mechanism. We showed
that standard upper bounds which are independent of interactions and order parameter
symmetry are only useful at extremely low carrier density and pratically useless anywhere
else. Our results have important implications on measurements of $T_c$ for
one band superconductors/superfluids with non-retarded interactions,
showing that any measurements that exceed $T_c^{\theta}$ must arise from
a non-BKT mechanism.

\acknowledgments{We thank the National Key R$\&$D Program of China (Grant 2018YFA0306501),
the National Natural Science Foundation of China (Grants 11522436 $\&$ 11774425),
the Beijing Natural Science Foundation (Grant Z180013), and the Research Funds of
Renmin University of China (Grants 16XNLQ03 $\&$18XNLQ15) for financial support.}



\begin{thebibliography}{2}
\bibitem{footnote-0}
To illustrate the point with an analogy, consider the distance between the Earth
and the Moon,  which is certainly smaller than the distance between the Earth and the Sun.
However, using the Earth-Sun distance as an upper bound for the Earth-Moon
distance is certainly an overkill, as the Moon apogee is the least upper bound (supremum)
for the Earth-Moon distance
\end{thebibliography}
\end{document}